\documentclass[letterpaper,twocolumn,superscriptaddress,floatfix]{revtex4}

\usepackage[latin1]{inputenc}
\usepackage{bm}
\usepackage{multirow,amssymb,amsbsy,amsmath}
\usepackage{graphicx}
\usepackage{verbatim}
\usepackage{braket}
\makeatletter
\usepackage{pifont}
\makeatother

\usepackage{graphicx}% Include figure files
\usepackage{dcolumn}% Align table columns on decimal point
\usepackage{bm}% bold math
\usepackage{ifthen}
\usepackage{booktabs}
\usepackage{nicefrac}
\usepackage{float}
\usepackage{color,xcolor}

\newcommand{\upcite}[1]{\textsuperscript{\textsuperscript{\cite{#1}}}}

\begin{document}
\renewcommand{\figurename}{Figure}
	
	\title{Generation of Spin Defects by Ion Implantation in Hexagonal Boron Nitride}
	\author{Nai-Jie Guo, Wei Liu,\textsuperscript{*} Zhi-Peng Li, Yuan-Ze Yang, Shang Yu, Yu Meng, Zhao-An Wang, Xiao-Dong Zeng, Fei-Fei Yan, Qiang Li, Jun-Feng Wang, Jin-Shi Xu, Yi-Tao Wang,\textsuperscript{*} Jian-Shun Tang,\textsuperscript{*} Chuan-Feng Li,\textsuperscript{*} and Guang-Can Guo}
	
	\date{\today }
	
	\begin{abstract}
		\textbf{Abstract:} Optically addressable spin defects in wide band gap semiconductors as promising systems for quantum information and sensing applications have recently attracted increasing attention. Spin defects in two-dimensional materials are expected to show superiority in quantum sensing due to their atomic thickness. Here, we demonstrate that an ensemble of negatively charged boron vacancy (V$ _\text{B}^{-} $) with good spin properties in hexagonal boron nitride can be generated by ion implantation. We carry out optically detected magnetic resonance measurements at room temperature to characterize the spin properties of ensembles of V$ _\text{B}^{-} $ defects, showing a zero-field splitting frequency of $ \sim $ 3.47 GHz. We compare the photoluminescence intensity and spin properties of V$ _\text{B}^{-} $ defects generated using different implantation parameters, such as fluence, energy and ion species. With the use of proper parameters, we can successfully create V$ _\text{B}^{-} $ defects with high probability. Our results provide a simple and practicable method to create spin defects in hBN, which is of great significance for realizing integrated hBN-based devices.
		
		%\centering
	\end{abstract}
	
	
	\maketitle %\nopagebreak
	\bibliographystyle{prsty}
	
	\section{INTRODUCTION}
	Solid-state spin defects have attracted widespread attention as promising quantum systems in recent decades \upcite{awschalom2013quantum} and have numerous applications in quantum information \upcite{togan2010quantum,waldherr2014quantum} and quantum sensing \upcite{maze2008nanoscale,kolkowitz2012coherent}. Some prominent systems have been studied extensively, including the nitrogen vacancy (NV) center \upcite{gali2019ab,plakhotnik2014all,stanwix2010coherence,wang2016coherence} and the silicon vacancy center \upcite{becker2018all,rogers2014all} in diamond and the divacancy center \upcite{li2020room,seo2016quantum} and silicon vacancy center \upcite{dong2019spin,carter2015spin} in silicon carbide. Although these defects have many remarkable properties, such as a long spin coherence time at room temperature \upcite{balasubramanian2009ultralong}, there are some intrinsic limitations due to the three-dimensional nature of the materials. For example, it is difficult to prepare spin defects close to the sample surface, which affects the sensitivity of the sensor \upcite{zhang2017depth}.
	
	Recently, the emergence of spin defects in two-dimensional materials and van der Waals crystals has provided a remedy for the limitations of three-dimensional materials. One of the outstanding materials is hexagonal boron nitride (hBN), which possesses a wide bandgap and a variety of atom-like defects, making hBN a good quantum system for single-photon emitters \upcite{xia2019room,tran2016robust,li2019purification,camphausen2020observation,bourrellier2016bright,mendelson2021identifying,Stern2021} and spin-addressable systems \upcite{mendelson2021identifying,Stern2021,gottscholl2020initialization,gottscholl2021room,liu2021rabi} at room temperature. Currently, most studies of spin defects are focused on the negatively charged boron vacancy (V$ _\text{B}^{-} $) that consists of a missing boron atom replaced by an extra electron in the hBN crystal \upcite{gottscholl2020initialization,kianinia2020generation,gao2020femtosecond,gottscholl2021room,liu2021rabi,liu2021temperature,gottscholl2021sub,murzakhanov2021creation,abdi2018color,Ivady2020,Sajid2020}. The V$ _\text{B}^{-} $ defects are photostable and exhibit good spin properties at room temperature \upcite{gottscholl2020initialization}. In addition, the V$ _\text{B}^{-} $ defects have a triplet ground state (S = 1) and can be initialized, manipulated and optically read out at room temperature, showing the potential for spin-based quantum information and sensing applications \upcite{gottscholl2021room,gottscholl2020initialization}.
	
	\begin{figure*}[htbp]
	    \centering
	    \includegraphics[width=0.75\linewidth]{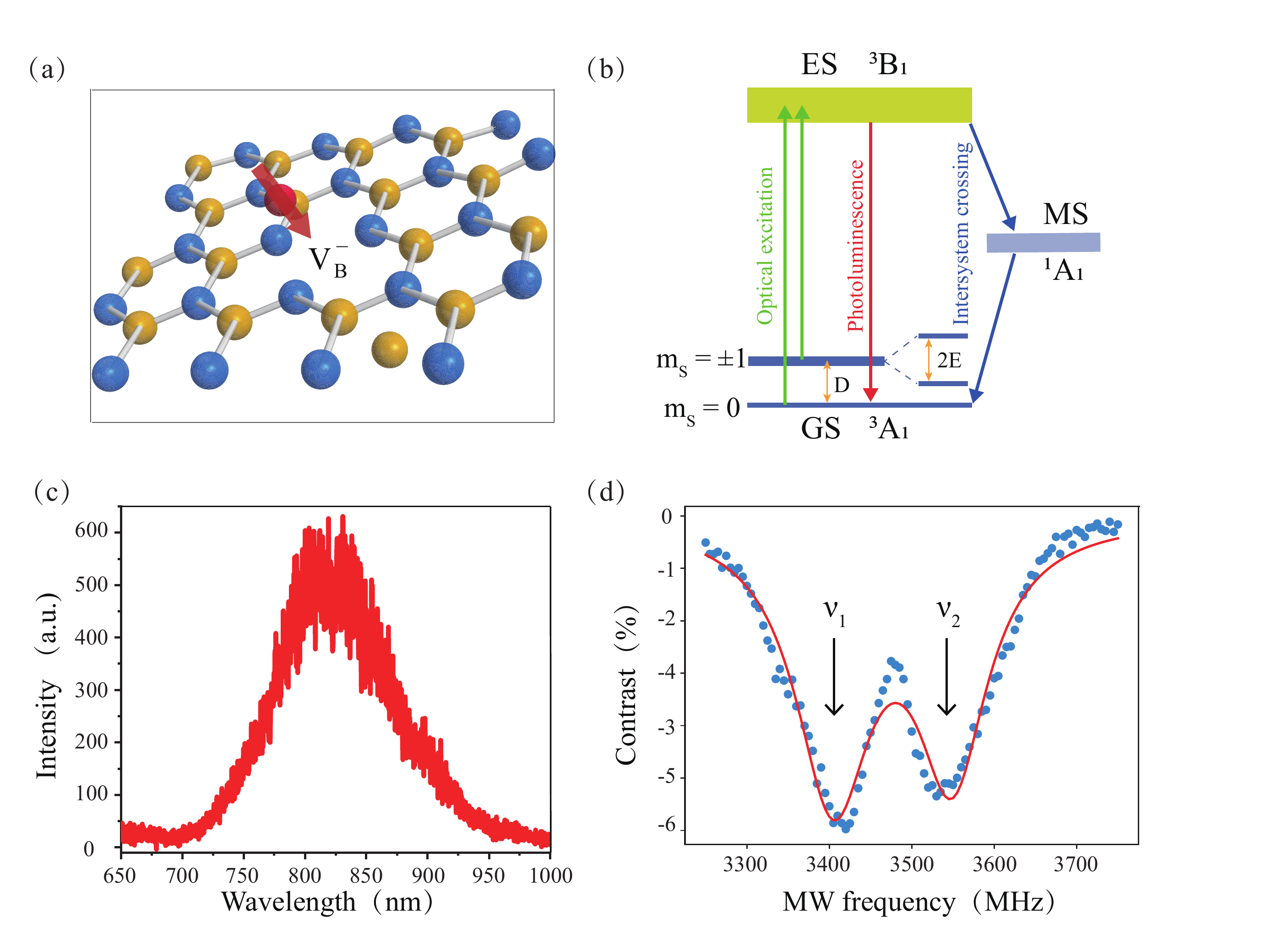}
	    \caption{Generation of V$ _\text{B}^{-} $ defects in hBN by implanting nitrogen ions with an energy of 30 keV and a fluence of 1 $ \times $ 10$ ^\text{14} $ ions/cm$ ^\text{2} $. (a) Schematic of the ion implantation process. Alternating boron (red) and nitrogen (blue) atoms form the crystalline hexagonal structure of an hBN monolayer. Implanted nitrogen (green) ions knock out boron atoms from the hBN lattice to generate V$ _\text{B}^{-} $ defects. (b) Simplified V$ _\text{B}^{-} $ energy-level diagram and the transitions among the ground state ($ ^3 A_1 $), excited state ($ ^3 B_1 $) and metastable state ($ ^1 A_1 $). (c) Photoluminescence (PL) spectrum for the implanted sample at room temperature, showing an emission centered at $\sim$ 820 nm. (d) ODMR measurement of the spin defects generated by ion implantation without an external magnetic field. The red line is a fit to a two-Lorentzian function, where $ \nu_{1} \sim $ 3405 MHz and $ \nu_{2} \sim $ 3548 MHz.}
	    \label{Figure 1}
	\end{figure*}
	
	In this context, we demonstrate a new way to generate V$ _\text{B}^{-} $ defects in hBN crystals by an ion implantation process using an ion implanter. At present, V$ _\text{B}^{-} $ defects can be generated by high-dose neutron irradiation \upcite{gottscholl2020initialization}, focused ion beam (FIB) implantation \upcite{kianinia2020generation}, femtosecond laser writing \upcite{gao2020femtosecond}, and high-energy electron irradiation \upcite{murzakhanov2021creation}. With appropriate energy and fluence for implanted ions, we successfully created ensembles of V$ _\text{B}^{-} $ defects using an ion implanter, which exhibit good contrast in optically detected magnetic resonance (ODMR) results. In addition, we measured the Rabi oscillations and spin-lattice relaxation time ($T _{1} $) of the defects (see Supporting Information).
	
	In the experiment, we used a commercially available monocrystalline hBN sample purchased from HQ Graphene with a lateral size of $ \sim $ 1 mm. Monocrystalline hBN is exfoliated with tape into 10$ \sim $100 nm-thick flakes, which are later transferred onto a silicon substrate. The sample is then put into an ion implanter (IonImplantatation-CETC-M56100) and the hBN flakes are implanted with parallelized ion beams over a large area. Through the ion implantation process we successfully created V$ _\text{B}^{-} $ defects. The process is schematically shown in Figure 1a. The high-energy ions break the B-N bonds in the hBN lattice and knock out boron atoms, leaving behind negatively charged vacancies. The photoluminescence (PL) and spin properties of the defects were characterized by using a confocal microscope system combined with a microwave system. We used a 532-nm laser to excite the defects with a laser power of 4.7 mW, a 0.5 N.A. objective (Olympus) to focus onto the sample and collect the fluorescence utilizing a 9-$ \mu $m-core-diameter fiber attached to an avalanche photodiode, and a copper wire with a diameter of 20 $ \mu $m placed close to the implanted sample as an antenna to deliver a microwave field \upcite{liu2021temperature,wang2019demand}.

	\section{RESULTS AND DISCUSSION}
	With the setup described above, we first characterized the PL spectrum of an ensemble of defects generated by implanting nitrogen ions with an energy of 30 keV and a fluence of 1 $ \times $ 10$ ^\text{14} $ ions/cm$ ^\text{2} $, as shown in Figure 1c. The implanted samples exhibit strong PL emission ranging from 700 nm to 1000 nm and a center at approximately 820 nm, which is the characteristic of V$ _\text{B}^{-} $ centers and consistent with reported V$ _\text{B}^{-} $ defects created by neutron irradiation, FIB and laser writing \upcite{gottscholl2020initialization,kianinia2020generation,gao2020femtosecond}. In addition, the V$ _\text{B}^{-} $ defects that we create are stable for an extented period of time at room temperature (see Supporting Information).
	
	\begin{figure*}[htbp]
	    \centering
	    \includegraphics[width=0.75\linewidth]{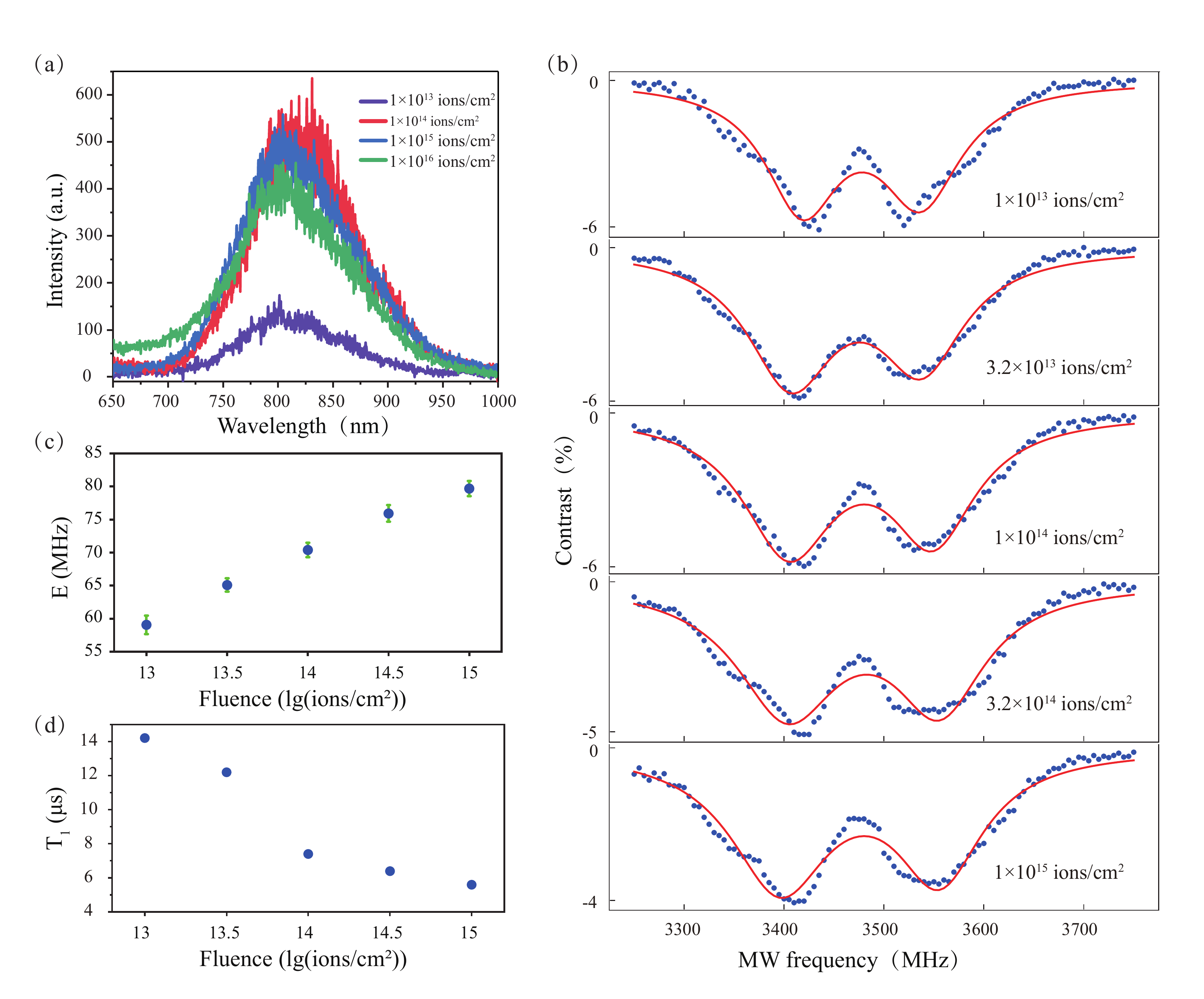}
	    \caption{Effects of implantation fluence on the defects. The fluence was varied from 1 $ \times $ 10$ ^\text{13} $ to 1 $ \times $ 10$ ^\text{16} $ ions/cm$ ^\text{2} $. The energy of the implanted nitrogen ions is fixed at 30 keV. (a) PL spectrum at room temperature for the defects created with different fluences. (b) ODMR measurements without an external magnetic field for the defects created with different fluences. (c) The ZFS parameter $ E $ as a function of fluence from 1 $ \times $ 10$ ^\text{13} $ to 1 $ \times $ 10$ ^\text{15} $ ions/cm$ ^\text{2} $. (d) The spin-lattice relaxation time $T _{1}$ as a function of fluence from 1 $ \times $ 10$ ^\text{13} $ to 1 $ \times $ 10$ ^\text{15} $ ions/cm$ ^\text{2} $.}
	    \label{Figure 2}
	\end{figure*}
	
	To further verify that the ensemble of defects generated by ion implantation are V$ _\text{B}^{-} $ centers, we performed optically detected magnetic resonance (ODMR) measurements at room temperature. ODMR measurements were carried out by scanning the frequency of the microwave field from 3250 MHz to 3750 MHz without an external magnetic field and the ODMR spectrum was fitted by a two-Lorentzian function, as shown in Figure 1d. The result indicates that the fluorescence signal drops when the microwave field oscillates at $ \nu_{1} \sim $ 3405 MHz and $ \nu_{2} \sim $ 3548 MHz, which is consistent with the ODMR spectra measured for V$ _\text{B}^{-} $ defects in previous works \upcite{gottscholl2020initialization,liu2021temperature}. Figure 1b shows that the $ m_{s}=\pm 1 $ excited state of the V$ _\text{B}^{-} $ center is more likely to return to the $ m_{s}=0 $ ground state through nonradiative intersystem crossing, so the V$ _\text{B}^{-} $ spin will be polarized into the $ m_{s}=0 $ ground state under continuous laser excitation. When the microwave frequency is in resonance with the split between the ground state sublevels, electrons in the $ m_{s}=0 $ state will be pumped into the $ m_{s}=\pm 1 $ state, leading to a decrease in fluorescence intensity \upcite{liu2021temperature}. The V$ _\text{B}^{-} $ center has a triplet ground state ($ S=1 $) with a zero-field splitting (ZFS) described by the parameters $ D $ and $ E $. The resonance frequencies $ \nu_{1} $ and $ \nu_{2} $ in the ODMR spectrum can be represented by $ \nu_{2,1}=D/h \pm \sqrt{E^{2}+(g\mu_{B}B)^{2}} $, where $ h $ is the Planck constant, $ g $ is the Land$ \acute{e} $ factor, $ \mu_{B} $ is the Bohr magneton and $ B $ is the static magnetic field \cite{gottscholl2020initialization}. In the absence of external magnetic field, the ZFS parameters $ D $ and $ E $ are given by $ D/h=(\nu_{1}+\nu_{2})/2 $ and $ E/h=(\nu_{2}-\nu_{1})/2 $. In our experiment, we find $ D/h=3475\pm 5 $ MHz and $ E/h=70\pm 5 $ MHz. The V$ _\text{B}^{-} $ defects exhibit a good ODMR contrast (up to 22\%, see Supporting Information) and a long relaxation time (up to 17 $ \mu $s, see Supporting Information) at room temperature, which shows promising spin properties for V$ _\text{B}^{-} $ defects generated by ion implantation.
	
	\begin{figure*}[htbp]
		\centering
		\includegraphics[width=0.75\linewidth]{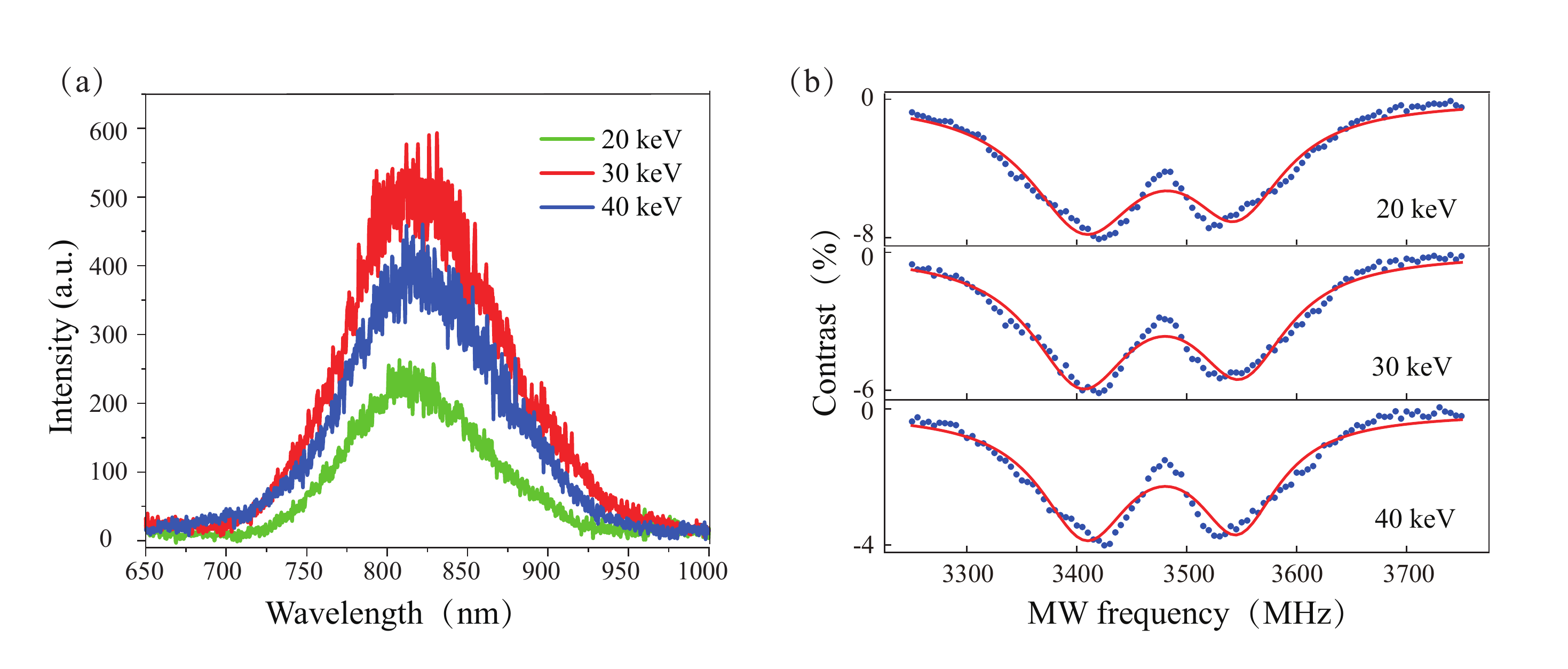}
		\caption{Effects of the energy of the implanted nitrogen ions on the defects. The energy is varied from 20 keV to 40 keV. The implantation fluence is fixed at 1 $ \times $ 10$ ^\text{14} $ ions/cm$ ^\text{2} $. (a) PL spectrum at room temperature for the defects created with different energies. (b) ODMR measurements without an external magnetic field for the defects created with different energies.}
		\label{Figure 3}
	\end{figure*}
	
	\begin{figure*}[htbp]
		\centering
		\includegraphics[width=0.75\linewidth]{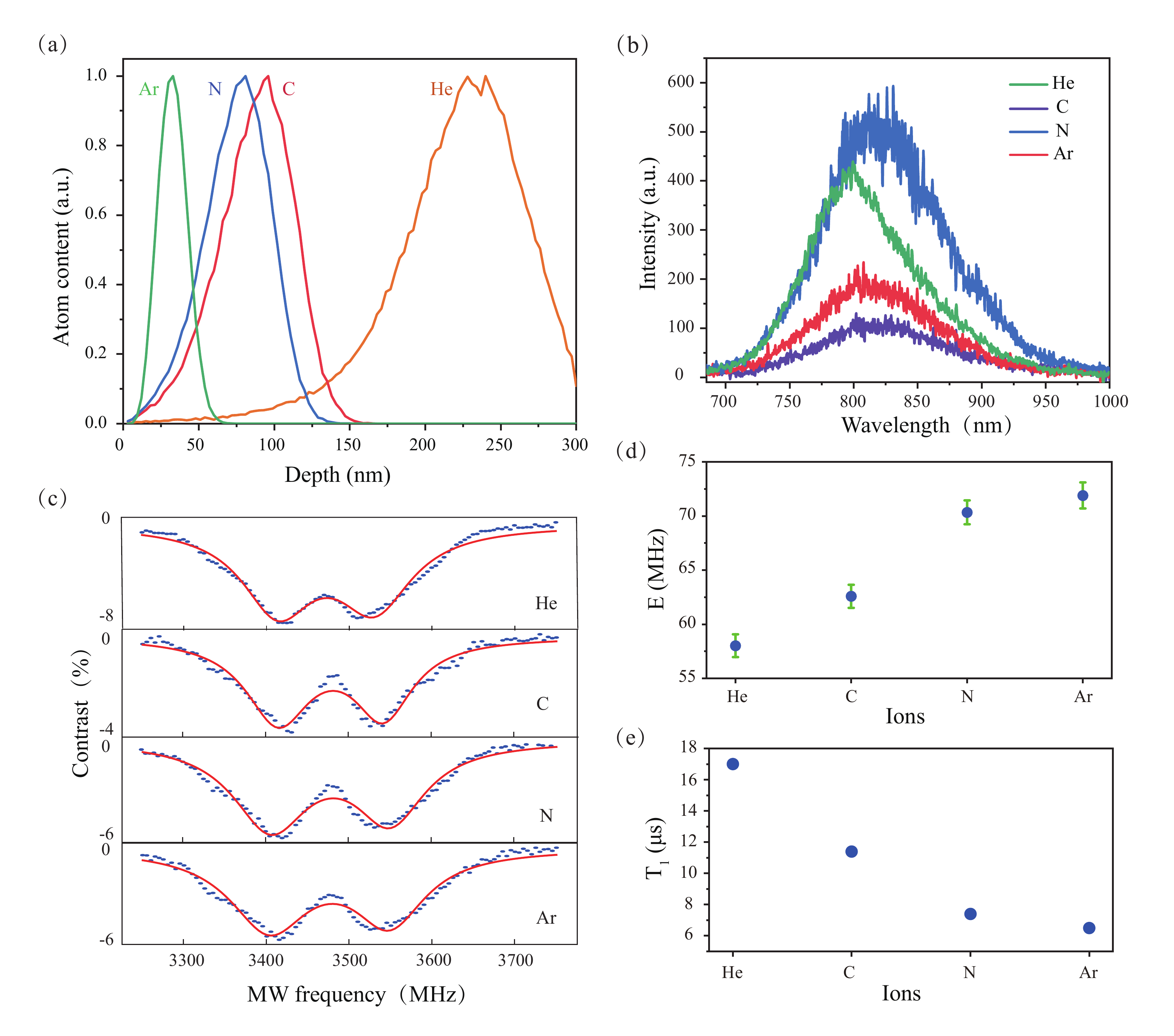}
		\caption{Effects of different implanted ion species on the defects. The implantation fluence was fixed at 1 $ \times $ 10$ ^\text{14} $ ions/cm$ ^\text{2} $ and the energy was fixed at 30 keV. (a) SRIM simulation of the defect distribution with depth generated by implanting different ions (He, C, N and Ar). (b) PL spectra at room temperature for the defects created with different ions. (c) ODMR measurements without an external magnetic field for the defects created with different ions. (d) The ZFS parameter $ E $ varies with different ions. (e) The spin-lattice relaxation time $T _{1}$ varies with different ions.}
		\label{Figure 4}
	\end{figure*}
	
	Next, we studied the effect of different implantation parameters, such as implantation fluence, energy and ion species. First, we compared the implantation effect of different fluences. We generated defects by implanting nitrogen ions with the same energy (30 keV) and increasing fluence from 1 $ \times $ 10$ ^\text{13} $ to 1 $ \times $ 10$ ^\text{16} $ ions/cm$ ^\text{2} $. Figure 2a shows a comparison of the room temperature PL spectra for four defect samples. We find that the intensity of the PL spectra increases with increasing fluence at low doses. When the fluence is increased up to 1 $ \times $ 10$ ^\text{14} $ ions/cm$ ^\text{2} $, the PL intensity is shown to decrease slightly, which is similar to that observed for V$ _\text{Si} $ defects in silicon carbide and NV and SiV centers in diamond \upcite{wang2019demand,schwartz2011situ,schroder2017scalable}. This decrease can be considered a saturation phenomenon for the V$ _\text{B}^{-} $ defects generated by ion implantation, which might result from the ion-induced damage of the crystal lattices that accumulates in the form of multiple vacancy defects \upcite{wang2019demand,schwartz2011situ,schroder2017scalable}. The ODMR spectra with two-Lorentzian fitting at different fluences are shown in Figure 2b. The measurements were carried out without an external magnetic field at room temperature. We find that the ZFS parameter $ D $ is stable at $ \sim $ 3475 MHz (see Supporting Information), while the ZFS parameter $ E $ increases almost linearly with the fluence ranging from 1 $ \times $ 10$ ^\text{13} $ to 1 $ \times $ 10 $ ^\text{15} $ ions/cm$ ^\text{2} $, as shown in Figure 2c. Nevertheless, when the fluence reaches to 1 $ \times $ 10$ ^\text{16} $ ions/cm$ ^\text{2} $, the ZFS parameter $ D $ is no longer stable and varies from 3460 MHz to 3520 MHz, and $ E $ is no longer linear (see Supporting Information). Furthermore, we measured the spin-lattice relaxation times $T _{1}$ of the defects, as shown in Figure 2d, which was found to be negatively correlated with the implantation fluence. The dependence of $ E $ and $T _{1}$ on fluence can be attributed to the increasing crystal damage with fluence. Damage due to ion implantation can give rise to local strain fields \upcite{vanDam2019Optical}, which has an effect on electron spin transitions. The strain field in hBN is mainly manifested as transverse strain when the damage is not very large \upcite{Jin2009Fabrication,Feng2018Imaging}, and the effect of transverse strain, as stated in Ref. \upcite{Teissier2014Strain}, is equivalent to a modification of $ E $. The more severe the damage, the larger the transverse strain, and then the larger the transverse splitting. Meanwhile, the damage can deteriorate the spin and optical coherence properties of defects \upcite{Tetienne2018Spin}, which suggests that more severe damage will lead to a shorter $T _{1}$.

	Then, to compare the implantation effect of different energies, we generated defects by implanting nitrogen ions with the same fluence (1 $ \times $ 10$ ^\text{14} $ ions/cm$ ^\text{2} $) and varying energy from 20 keV to 40 keV. Figure 3a shows a comparison of the PL spectra measured for these defect samples, and Figure 3b shows a comparison of the ODMR spectra measured at room temperature. We can see that different energies mainly affect the PL intensity but have almost no effect on the spin properties of the V$ _\text{B}^{-} $ defects. The PL spectrum for the V$ _\text{B}^{-} $ defects generated at 30-keV energy displays higher intensity than that of V$ _\text{B}^{-} $ defects generated at other energies, while the ODMR spectra of the V$ _\text{B}^{-} $ defects generated at different energies display the same resonance frequency, i.e., the same ZFS parameters $ D $ and $ E $. Additionally, we find that the spin lattice relaxation times $T _{1}$ scarcely change with the implantation energies (see Supporting Information). Because our hBN samples are 10$ \sim $100 nm-thick flakes, ions can easily penetrate the flakes rather than remain in the samples, even at low energy. Although the implantation energies are different, the damage due to ion implantation with the same ion species and fluence is similar. Therefore, implantation energy does not affect the spin properties of the defects.
	
	Finally, to compare the implantation effect of different ion species, we generated defects by implanting nitrogen, argon, helium and carbon ions with the same fluence (1 $ \times $ 10$ ^\text{14} $ ions/cm$ ^\text{2} $) and energy (30 keV), respectively. We simulate the theoretical distribution of the V$ _\text{B}^{-} $ defects with depth created by different ions using a stopping-and-range-of-ions-in-matter (SRIM) simulation, as shown in Figure 4a. The result indicates that the number of generated defects and the penetration depth are obviously different for different ions. Argon ions are more likely to create shallow defects, while helium ions can be used to create defects in thicker samples. Figure 4b shows a comparison of the PL spectra measured for these samples at room temperature, from which we can see that the sample implanted using nitrogen ions has the highest PL intensity. Figure 4c shows the ODMR spectra for the defects generated by different implantation ions, indicating that they all have spin properties. We find that when we implant different ions, the ZFS parameter $ D $ is stable at $ \sim $ 3475 MHz (see Supporting Information), while the ZFS parameter $ E $ is different, which is similar to implantation with different fluences. The difference in the ZFS parameter $ E $ is shown in Figure 4d, showing that the ZFS parameter $ E $ increases as the ion radius increases. Additionally, the spin-lattice relaxation times $T _{1}$ for the defects decrease as the ion radius increases, as shown in Figure 4e. Similar to the dose-dependent relationship mentioned above, the dependence of $ E $ and $T _{1}$ on the ion species can also be attributed to crystal damage. With increasing atomic number, the damage increases due to the larger collision cross-section, thus, $ E $ increases and $T _{1}$ decreases.
	
	\section{CONCLUSIONS}
	We successfully generated optically active V$ _\text{B}^{-} $ defects by ion implantation in hBN. There are also several other ways to generate V$ _\text{B}^{-} $ defects, such as the neutron-irritation method, the FIB method, the laser-writing method, and the electron-irradiation method. All these methods have their own advantages and disadvantages. For example, the neutron-irritation method is the primary way to generate V$ _\text{B}^{-} $ defects, but it needs to be carried out in a nuclear reactor, which is not very convenient and slightly expensive. Comparatively, the ion-implantation method is convenient and inexpensive because the ion implanter is commercially available. The FIB method allows for the patterning of arrays of spin defects due to its controllability and good positioning. This is the advantage of this method. However, to the best of our knowledge, the ion source often used for commercial FIBs is Ga, and the use of other ion sources is rare. In contrast, the ion implanter has a variety of available ion sources (He$ ^{+} $, C$ ^{+} $, N$ ^{+} $, Ar$ ^{+} $, etc.), and, moreover, the ion-implantation method can be used to create V$ _\text{B}^{-} $ defects over a large scale. The laser-writing method is simple and flexible, as it can be conducted in an ambient environment with no vacuum requirement. However, the required femtosecond laser pulse has a relatively large energy and is possibly destructive towards the sample. In comparison, the ion-implantation method is gentle and results in little damage to the sample. In addition, the electron irradiation method makes it possible to avoid clustering of defects, but it requires very high electron energy (2 MeV). Relatively, the ion-implantation method needs only a relatively low ion energy (30 keV). Therefore, our ion-implantation method will be a good supplement to all the above-mentioned methods.
	
	Our results show that the implantation parameters, such as fluence, energy and ion species, have clear effects on the PL intensity and spin properties of ion-implantation-generated V$ _\text{B}^{-} $ defects. Therefore, we can create good ensembles of V$ _\text{B}^{-} $ defects with high probability by adjusting the fluence (1 $ \times $ 10$ ^\text{14} $ ions/cm$ ^\text{2} $) and energy (30 keV) of the implanted nitrogen ions, and the V$ _\text{B}^{-} $ defects exhibit a good ODMR contrast at room temperature, which is important for spin-addressable systems. Furthermore, we find that the defects created by implanting helium ions with an energy of 30 keV and a fluence of 1 $ \times $ 10$ ^\text{14} $ ions/cm$ ^\text{2} $ have the longest spin-lattice relaxation time of 17 microseconds at room temperature, which is comparable to that achieved for defects created by neutron irradiation \upcite{liu2021rabi,gottscholl2021room}. Our work provides a simple and practicable method for controllable engineering of spin defects in hBN and paves the way for integrated quantum information and sensing applications. 
	
	\section*{Supporting Information}
	This material is available free of charge via the internet at xxxxxxxxxxxx.
	
	Additional data, PL stability, and probablity of generating a V$ _\text{B}^{-} $ center.
	
	\section*{Author Information}
	
	\subsection*{Corresponding Authors}
	
	\textbf{Wei Liu} - CAS Key Laboratory of Quantum Information and CAS Center For Excellence in Quantum Information and Quantum Physics, University of Science and Technology of China, Hefei 230052, People's Republic of China; Email: lw691225@ustc.edu.cn
	
	\textbf{Yi-Tao Wang} - CAS Key Laboratory of Quantum Information and CAS Center For Excellence in Quantum Information and Quantum Physics, University of Science and Technology of China, Hefei 230052, People's Republic of China; Email: yitao@ustc.edu.cn
	
	\textbf{Jian-Shun Tang} - CAS Key Laboratory of Quantum Information and CAS Center For Excellence in Quantum Information and Quantum Physics, University of Science and Technology of China, Hefei 230052, People's Republic of China; Email: tjs@ustc.edu.cn
	
	\textbf{Chuan-Feng Li} - CAS Key Laboratory of Quantum Information and CAS Center For Excellence in Quantum Information and Quantum Physics, University of Science and Technology of China, Hefei 230052, People's Republic of China; orcid.org/0000-0001-6815-8929; Email: cfli@ ustc.edu.cn
	
	\subsection*{Authors}
	\textbf{Nai-Jie Guo} - CAS Key Laboratory of Quantum Information and CAS Center For Excellence in Quantum Information and Quantum Physics, University of Science and Technology of China, Hefei 230052, People's Republic of China; Email: guonaijie@mail.ustc.edu.cn
	
	\textbf{Zhi-Peng Li} - CAS Key Laboratory of Quantum Information and CAS Center For Excellence in Quantum Information and Quantum Physics, University of Science and Technology of China, Hefei 230052, People's Republic of China; Email: lzp0116@mail.ustc.edu.cn
	
	\textbf{Yuan-Ze Yang} - CAS Key Laboratory of Quantum Information and CAS Center For Excellence in Quantum Information and Quantum Physics, University of Science and Technology of China, Hefei 230052, People's Republic of China; Email: yyz14@mail.ustc.edu.cn
	
	\textbf{Shang Yu} - CAS Key Laboratory of Quantum Information and CAS Center For Excellence in Quantum Information and Quantum Physics, University of Science and Technology of China, Hefei 230052, People's Republic of China; Email: yushang@mail.ustc.edu.cn
	
	\textbf{Yu Meng} - CAS Key Laboratory of Quantum Information and CAS Center For Excellence in Quantum Information and Quantum Physics, University of Science and Technology of China, Hefei 230052, People's Republic of China; Email: mengyu23@mail.ustc.edu.cn
	
	\textbf{Zhao-An Wang} - CAS Key Laboratory of Quantum Information and CAS Center For Excellence in Quantum Information and Quantum Physics, University of Science and Technology of China, Hefei 230052, People's Republic of China; Email: zawang@mail.ustc.edu.cn
	
	\textbf{Xiao-Dong Zeng} - CAS Key Laboratory of Quantum Information and CAS Center For Excellence in Quantum Information and Quantum Physics, University of Science and Technology of China, Hefei 230052, People's Republic of China; Email: zengxiaodong@mail.ustc.edu.cn
	
	\textbf{Fei-Fei Yan} - CAS Key Laboratory of Quantum Information and CAS Center For Excellence in Quantum Information and Quantum Physics, University of Science and Technology of China, Hefei 230052, People's Republic of China; Email: yanff@mail.ustc.edu.cn
	
	\textbf{Qiang Li} - CAS Key Laboratory of Quantum Information and CAS Center For Excellence in Quantum Information and Quantum Physics, University of Science and Technology of China, Hefei 230052, People's Republic of China; Email: qianglee@ustc.edu.cn
	
	\textbf{Jun-Feng Wang} - CAS Key Laboratory of Quantum Information and CAS Center For Excellence in Quantum Information and Quantum Physics, University of Science and Technology of China, Hefei 230052, People's Republic of China; Email: jfwang89@ustc.edu.cn
	
	\textbf{Jin-Shi Xu} - CAS Key Laboratory of Quantum Information and CAS Center For Excellence in Quantum Information and Quantum Physics, University of Science and Technology of China, Hefei 230052, People's Republic of China; Email: jsxu@ustc.edu.cn
	
	\textbf{Guang-Can Guo} - CAS Key Laboratory of Quantum Information and CAS Center For Excellence in Quantum Information and Quantum Physics, University of Science and Technology of China, Hefei 230052, People's Republic of China; Email: gcguo@ustc.edu.cn

	\section*{ACKNOWLEDGMENTS}
	This work was supported by the National Key Research and Development Program of China (No. 2017YFA0304100), the National Natural Science Foundation of China (Grants Nos. 12174370, 11822408, 12174376, 11774335, 11821404, and 11904356), the Open Research Projects of Zhejiang Lab (NO. 2021MB0AB02), the Key Research Program of Frontier Sciences of the Chinese Academy of Sciences (Grant No. QYZDY-SSW-SLH003), the Fok Ying-Tong Education Foundation (No. 171007), Science Foundation of the CAS (No. ZDRW-XH-2019-1), Anhui Initiative in Quantum Information Technologies (AHY020100, AHY060300), the Fundamental Research Funds for the Central Universities (Nos. WK2470000026, WK2030000008 and WK2470000028). This work was partially carried out at the USTC Center for Micro and Nanoscale Research and Fabrication.

\end{document}